# Localization and electron-electron interaction effects in magnetoresistance of *p*-type Ge/Ge$_{1-x}$Si$_x$ heterostructures

Yu.G. Arapov[1], G.I. Harus, V.N. Neverov, A.T. Lonchakov,
N.G. Shelushinina and M.V. Yakunin

*Institute of Metal Physics, Ekaterinburg, 620219 Russia*

## 1. Introduction

The diffusive nature of electron motion in a disordered system results in quantum corrections to the Drude conductivity $\sigma_0$ due to both the single particle weak localization (WL) effects and the disorder-modified electron-electron interaction (EEI) [1]. The WL and EEI parts of the total quantum correction $\Delta\sigma$ may be separated by a magnetic field $B$ since the localization effect should be suppressed for $B >> B_{tr} = \hbar c/2e\ell^2$ ($\ell$ is the elastic free path) [2]. In contrast to the $B$ sensitivity of $\Delta\sigma$, for the EEI effect it is found that $\Delta\sigma_{xy} = 0$ and $\Delta\sigma_{xx} = \Delta\sigma_{ee}$ ($B = 0$) at both $\omega_c\tau < 1$ and $\omega_c\tau > 1$ [3]. As a consequence, at $B >> B_{tr}$ the following expression for longitudinal resistivity takes place [3,4]:

$$\rho_{xx}(B) = \rho_0 + \rho_0^2[1-(\mu B)^2]\Delta\sigma_{ee} \qquad (1)$$

with $\rho_0 = 1/\sigma_0$. The interplay of classical cyclotron motion and EEI effect thus leads to the parabolic negative magnetoresistance (MR) [4,5]:

$$\Delta\rho/\rho_0 = -(\mu B)^2 \Delta\sigma_{ee}/\sigma_0. \qquad (2)$$

## 1. Results and discussion

We report on the results of the MR investigations for two samples 1124$b_3$(1125$a_7$) of strained multilayer *p*-type Ge/Ge$_{1-x}$Si$_x$ heterostructures with hole densities $p = 2.5(2.8)\cdot 10^{11}$ cm$^{-2}$ and mobilities $\mu = 1(1.7)\cdot 10^4$ cm$^2$/Vs at $T \geq 0.1$ K in fields up to 2T. The logarithmic upturn of $B = 0$ resistivity takes place in both samples at $T < 20$ K (fig.1). For sample 1124$b_3$, the negative MR is observed in the whole range of magnetic fields up to $\mu B = 1$ ($B \approx 1$T) at $T < 12$ K (fig.2). The same situation takes place for sample 1125$a_7$ at $T < 1.3$ K. But at $T \geq 1.3$ K MR, also being negative for $B < 10B_{tr} \approx 0.2$ T, becomes positive for $B \geq 0.2$ T (Fig.3).

Estimation of the Fermi energy for complex Ge valence band in both samples yields $E_F \approx \Delta$, where $\Delta$ is the energy separation of the ground and the second hole subbands. Variation of internal uniaxial strain, as well as some differences in the Ge layer width, may result in variation of $\Delta$ and thus in the degree of the second subband filling. We connect the observed positive MR just with the interplay of two types of holes due to the partial

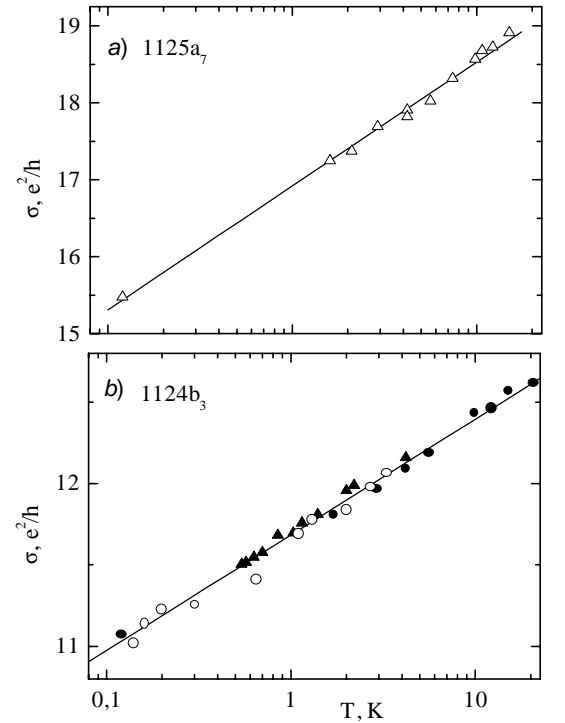

Fig.1. Temperature dependencies of conductivity for samples 1125$a_7$ and 1124$b_3$

---

[1] E-mail: arapov@imp.uran.ru

filling of the second subband.

For two types of carriers with conductivities $\sigma_2 \ll \sigma_1$ we obtain the modification of (2):

$$\Delta\rho/\rho_0 = (\mu B)^2 (\sigma_2 - \Delta\sigma_{ee})/\sigma_0. \qquad (3)$$

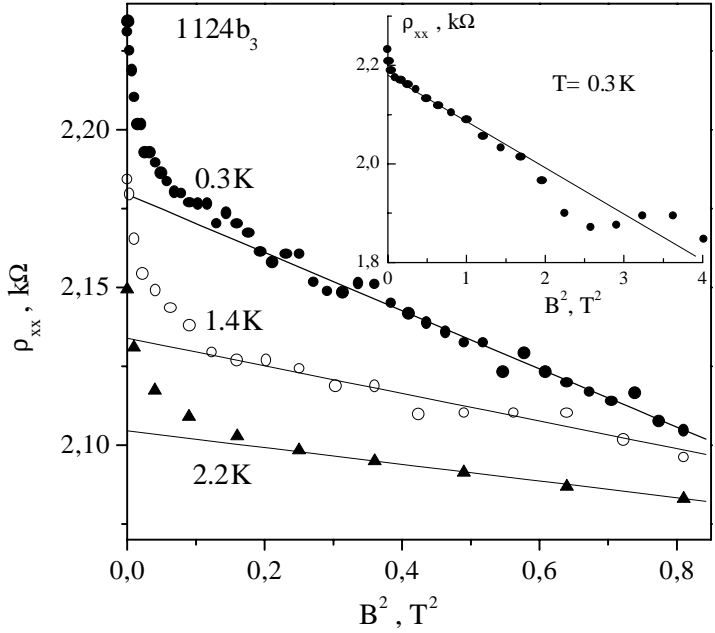

Fig.2. Resistivity $\rho_{xx}$ against $B^2$ for sample 1124$b_3$. Solid lines are the extrapolation of the $B^2$ dependence to zero field. The inset shows the curve for $T = 0.3$K at higher fields ($\omega_c \tau > 1$).

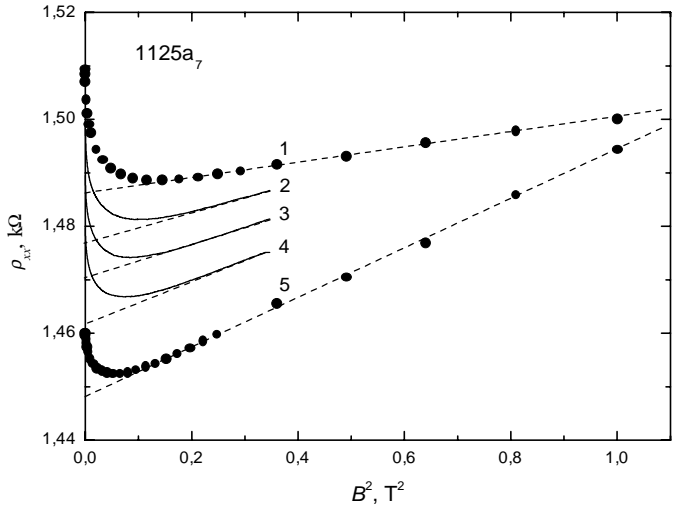

Fig.3. Resistivity $\rho_{xx}$ against $B^2$ for sample 1125$a_7$. $T$ [K] = 1 - 1.55, 2 - 2.0, 3 - 2.5, 4 – 3.0, 5 – 4.2.

As $\Delta\sigma_{ee}$ depends logarithmically on $T$, MR may be *negative* at sufficiently low $T$ ($\Delta\sigma_{ee} > \sigma_2$) and give place to the *positive* one with increasing $T$ ($\Delta\sigma_{ee} < \sigma_2$). Important is that for one or two types of carriers, as well as for negative or positive MR, extrapolation of the high-field ($B \gg B_{tr}$) $\rho_{xx}(B^2)$ dependence to $B = 0$ yields the same result:

$$\rho_{xx}(B \to 0) = \rho_0 (1 + \rho_0 \Delta\sigma_{ee}) \quad (4)$$

and makes it possible to estimate $\Delta\sigma_{ee}$ and thus to separate WL and EEI contributions to the total quantum correction at $B = 0$ (see figures 2,3). The result for both of our structures is that $\Delta\sigma_{ee} \approx 2/3 \Delta\sigma$ and $\Delta\sigma_{WL} \approx 1/3 \Delta\sigma$.

For *p*-type structures with strong spin-orbit coupling, *positive* weak-localization MR is predicted for situation with $E_F \approx \Delta$ due to a mixing of the heavy- and light-hole states [6-8]. Contrary to this prediction, it is seen from figures 2 and 3 that MR due to WL effect is *negative* at $B < 10 B_{tr}$ for both of our samples. The disparity of our data with the theory [8] may be due to the different kind of impurity scattering potential (remote impurities [9], but not the short-range scatterers) or to the unconsidered role of the second hole subband, when the analysis of data with due regard for a relation between the times of phase breaking and intersubband transitions is needed [7,8].

But the main reason we believe is the strong uniaxial tension strain of Ge layers taking place in our heterostructures [10]. Due to uniaxial tension, a set of heavy-hole subbands is shifted to lower energies while a set of light-hole subbands is shifted to higher energies (if the hole energy is counted upwards) [11]. As a result, the interplay of the lowest different-type hole subbands becomes weaker and the degree of heavy-hole/light-hole mixing at the Fermi level diminishes.

## 3. Conclusions

In summary, we analyze magnetoresistance data for *p*-type Ge/Ge$_{1-x}$Si$_x$ heterostructures with Fermi energy near the bottom of the second confinement hole subband ($E_F \cong \Delta$). The *negative* MR due to a suppression of WL effect in the weak-field ($B < B_{tr}$) region gives evidence that we are not in a situation with the strong spin-orbit scattering, in contrast with results of the theory [6,7]. Extrapolation of the $B^2$ dependencies of the high-field MR ($B >> B_{tr}$) to zero field is used to separate the WL and EEI parts of the total quantum correction to the conductivity.

## Acknowledgements

The work is supported in part by Russian Foundation for Basic Researches, projects 01-02-17685, 02-02-16401, 02-02-06864, 02-02-06168, INTAS YSF 01/1-156 and the 6-th concurs-expertise RAS (1999) No. 68.